# Positronium formation in graphene and graphite

V. A. Chirayath[1, a)], A. J. Fairchild[1], R. W. Gladen[1], M. D. Chrysler[1], A. R. Koymen[1] and A. H. Weiss[1]

[1]*Department of Physics, University of Texas at Arlington, Arlington, Texas – 76019*

a)*Corresponding author:* chirayat@uta.edu

**Abstract.** Positronium (Ps) formation on the surface of clean polycrystalline copper (Cu), highly oriented pyrolytic graphite (HOPG) and multi layer graphene (MLG) grown on a polycrystalline copper substrate has been investigated as a function of incident positron kinetic energy (1.5eV to 1keV). Measurments on Cu indicate that as the kinetic energy of the incident positrons increases from 1.5eV to 900eV, the fraction of positrons that form Ps ($f_{Ps}$) decreases from ~0.5 to ~0.3. However, in HOPG and MLG, instead of a monotonic decrease of $f_{Ps}$ with positron kinetic energy, a sharp peak is observed at ~ 5eV and above ~200eV, $f_{Ps}$ remains nearly constant in HOPG and MLG. We propose that in HOPG and MLG, at low incident positron energies the Ps formation is dominated either by a surface Plasmon assisted electron pick up process or by an energy dependent back scattering process. Both these processes can explain the peak observed and the present data can help to augment the understanding of Ps formation from layered materials.

## INTRODUCTION

Positronium formation depends on the surface electronic structure of the sample and can, in principle be used as a highly surface selective probe to study variations of the surface electronic structure [1]. Investigation of various channels of Ps formation may help to design surfaces that produce Ps with the high level efficiency that is a requirement for several planned experiments with high intensity Ps beam [2]. The physical processes leading to Ps formation on surfaces are still under active investigation [3]. For example, very recently a new method of spontaneous Ps emission from p-Si(100) surface was observed that was suggested to be due to an exotic positron-electron state which becomes accessible in the presence of positrons on the surface [4]. In this paper, we present first investigation of Ps formation from 6-8 layers of graphene (MLG) on a polycrystalline Cu substrate. The Ps formation from MLG was compared to that from HOPG and from a sample of clean polycrystalline Cu. Our results indicate that Ps formation on graphene and graphite is similar and is quite different from Ps formation on a Cu surface. The measured positronium yield, as a function of incident positron energy, peaks at ~5eV for MLG and HOPG. In contrast, in the case of Cu, the positronium yield slowly decreases and does not have any peak like structure. The results show that Ps formation from graphitic samples (MLG and HOPG) at low energies is dominated by excitation of surface Plasmon modes or (and) by electron pick-up during back scattering of positrons through surfaces.

## EXPERIMENTS

The experiments were carried out using the time of flight (TOF) positron annihilation induced Auger electron spectroscopy (PAES) system at the University of Texas at Arlington (Fig.1) [5, 6]. The positron beam utilizes a micrometer thick tungsten single crystal (W (100)) as a moderator in the transmission geometry. The beam optics allows for the transport of 1eV positrons from the moderator to the sample. The energy of the positrons on the sample can be varied by appropriately biasing the sample with respect to the moderator. The low energy positrons are bent around the electron detector using an ExB field. The positrons travel through a 1 m field-free TOF tube before being deposited on the sample surface. The electrons emitted from the sample surface as a result of positron implantation and annihilation travel through the TOF tube before getting bent into the electron detector by the E x B field. The TOF tube is also used to measure the energy of positrons implanted by using it as a retarding field analyser.

The annihilation gamma rays are detected using a 2 inch NaI (Tl) detector and a 2 inch BaF$_2$ detector. The fraction of positrons that form Ps ($f_{Ps}$) was calculated from the gamma spectra measured with NaI(Tl) detector using the method described by Mills et al. [7]. The timing signals from the BaF$_2$ detector and the electron detector were used to generate the TOF spectra of the Auger electrons emitted from the top atomic layer of the sample as a result of positron annihilation induced core or valence holes. The measurements were carried out on samples consisting of a freshly cleaved highly oriented pyrolytic graphite (HOPG, ZYA grade from SPI supplies), multiple layers of graphene (~ 6 – 8) grown on a polycrystalline Cu foil (MLG; from ACS materials) and the same surface after the removal of the graphene layer by sputtering (Cu).

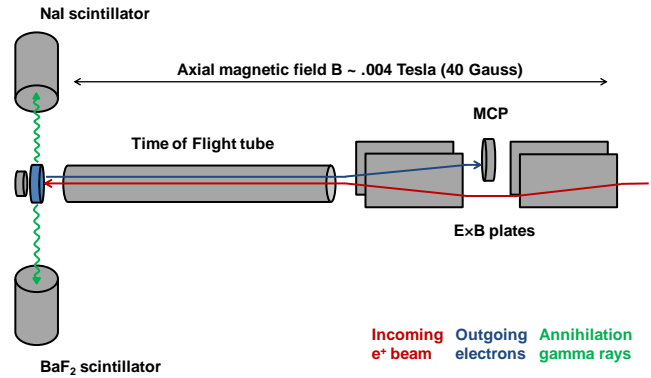

**FIGURE 1.** Apparatus used to obtain the Ps fraction and PAES from the surface of Cu, Cu with multilayer Graphene over-layers and HOPG. The positron beam has ~ 1 m long time of flight tube for the transport of positron annihilation induced Auger electrons to the electron detector. The electron and positron flight paths are separated using E x B plates placed near the electron detector. The annihilation gamma rays are detected using a BaF$_2$ detector and a NaI(Tl) scintillation detector. The timing signal between the BaF$_2$ detector and the electron detector is used to generate the TOF spectra of positron annihilation induced Auger electrons and the gamma spectra measured with NaI(Tl) detector was used to measure the Ps fraction.

## RESULTS AND DISCUSSION

Figure 2 (a) shows the variation of $f_{Ps}$ as a function of the incident positron kinetic energy (1.5eV to 0.9keV). Figure 2 (b) shows the same plot in an expanded energy scale to clarify the specific features seen in the $f_{Ps}$ vs. positron kinetic energy in HOPG and MLG at positron energies less than 80 eV. Some of the key differences between the Cu and graphite samples in the variation of $f_{Ps}$ with positron energy are pointed out below. The $f_{Ps}$ from Cu surface is greater than or equal to that from HOPG and MLG throughout the measured energy range. The $f_{Ps}$ decreases with increasing positron energy in Cu. However, for both HOPG and MLG there is a sharp peak at ~ 5eV and a broad shoulder at ~ 20eV. The $f_{Ps}$ is nearly constant in HOPG and MLG above ~ 200eV, whereas in Cu, $f_{Ps}$ shows a decreasing trend even at 0.9keV.

Sferlazzo et al. [8] had previously shown that $f_{Ps}$ in graphite is only about ~ 16% when measured with 300eV positrons. This was surprising as graphite was shown to have a negative Ps work function ($\varphi_{Ps} = -0.6eV$) which points to spontaneous Ps emission. Energy conservation allows all electrons within 0.6eV from the Fermi level to take part in Ps formation. The Ps formed with electron at the Fermi level will be emitted from the surface with the maximum energy equal to the absolute value of the Ps work function. However, the electrons near the Fermi level in graphite have high parallel momentum [9]. If parallel momentum is conserved during thermal Ps formation, any Ps formed by the electrons at the Fermi level must have energies larger than 0.6eV. Thus, Ps formation is forbidden by the simultaneous requirement of energy and parallel momentum conservation. Sferlazzo et al [8] using 2D angular correlation of annihilation radiation (ACAR) measurements and temperature dependence of Ps formation, showed that Ps can however be formed through a higher order process involving absorption or emission of phonons which was hypothesized to be responsible for the small Ps fraction observed at 300eV. On the other hand direct Ps formation on Cu surface is not forbidden which explains the relatively higher $f_{Ps}$ when compared to graphite. The $f_{Ps}$ for positron energies above 200eV is small (~ 20%) in both HOPG and MLG which is consistent with the reported values in the literature and it points to Ps formation through the phonon absorption/emission process by positrons thermalized in the bulk of graphite or graphene over layers. However, the $f_{Ps}$ is approximately a constant for positron energies above 200eV in HOPG and MLG. One would have expected to see a decrease in $f_{Ps}$ with increasing positron energies (as seen for Cu) due to the reduced probability of higher energy positrons to diffuse back to the surface and form Ps [7]. The constant values indicates that Ps formation may be as efficient in the inter layers in graphite (or MLG) [10] as at the outer most surface which may be due to the similarity of the electronic cloud sampled by positrons in the inter-layer region to the electron cloud sampled at the outermost surface.

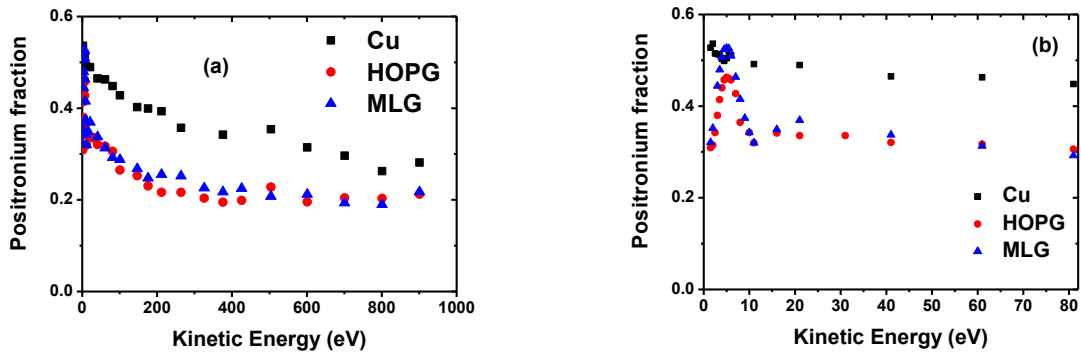

**FIGURE 2.** (a) The variation of $f_{Ps}$ with positron kinetic energy in Cu, HOPG and MLG. The $f_{Ps}$ from HOPG and MLG are less than that from Cu except at low positron kinetic energies. Above 200eV the $f_{Ps}$ from HOPG and MLG is constant whereas it continues to decrease in Cu. (b) shows the same data as in (a) but on an expanded scale to clearly see the features in HOPG and MLG. Both HOPG and MLG show a sharp peak at ~ 5eV and broad feature at ~ 20eV. Cu on the other hand shows a slow decrease with increasing positron energy.

As the positron energy is reduced to values below 200eV, the $f_{Ps}$ measured from HOPG and MLG increases above 20%. Since Ps formation by thermalized positrons can only result in ~ 20% Ps formation from graphite because of the constraints from energy and momentum conservation, an increase in $f_{Ps}$ points to Ps formation by positrons that can overcome the energy and parallel momentum conservation barrier. This is possible if the Ps is formed by non thermalized positrons during back scattering [11]. Previous studies by Howell et al. [12] have shown that at positron energies less than 100eV, Ps formation through inelastic backscattering through the surface may be more intense than the Ps formed by thermalized positrons. The Ps intensity by backscattering depends on the back scattering probability as well as on the electron pick up probability which depend on the incident positron energy. There are no direct measurements of the Ps formation probability by scattering through surface atoms as a function of positron energy. However, there are scattering studies from gases which has maximum Ps formation probability at low positron energies (~ 10 – 50eV) [12]. If we assume similar energy dependence for Ps formation through positron backscattering from surfaces atoms; the increase in the Ps intensity at positron energies less than 200eV for HOPG, MLG can be understood to be due to the increase in positron backscattering co-efficient with decreasing positron energy that overlaps with the increase in Ps formation probability at low positron energies.

The main characteristic of the variation of $f_{Ps}$ with positron energy from graphitic samples (HOPG and MLG) is the presence of the sharp peak at ~ 5eV. The presence of the peak was confirmed by following the intensity of the Auger electron peak from the samples (Cu, HOPG and MLG) as a function of incident positron energy. Positron annihilation induced Auger electron emission competes with the Ps formation at the surface. The Auger process is initiated by the annihilation of a core or valence electron with a positron that is trapped in the surface state. The hole, thus created by the annihilation, is filled by an electron at a higher energy state. The energy released as a result of this transition is transferred to a third electron which may then have enough energy to overcome the surface barrier. Hence, the detection of an Auger electron implies that the positron was not emitted from the surface as Ps and the detection of Ps points to the absence of positron annihilation from the surface state. The peak in the Ps fraction ($f_{Ps}$) observed for graphitic samples should therefore correspond to a dip in the emission of positron annihilation induced Auger electrons. Fig. 3 shows the fraction of positrons reaching the surface that initiate Auger emission. The fraction was calculated from the intensity of the 60eV core Auger peak that corresponds to an $M_{2,3}$VV Auger transition in Cu (Fig. 3(a)) and from the intensity of the 263eV KVV Auger transition in HOPG (Fig. 3(b)) and MLG (Fig. 3(c). The intensity of the Auger peaks follows the variation of the Ps intensity inversely that confirms the presence of the peak in the variation of $f_{Ps}$ with the incident positron energy seen in HOPG and MLG.

We propose two independent hypotheses to explain the observed peak in positronium formation from graphitic surfaces at the incident positron energy of 5eV. The Ps formed at low incident positron energies depend on the positron backscattering intensity and the electron capture probability. The peak at 5eV and the broad feature at 20eV we see in our measurements can correspond to the presence of similar features in the positron backscattering cross section and (or) in the Ps formation probability. For example, Gullikson and Mills [13] had shown that the intensity of elastically backscattered positrons from HOPG surface show appreciable reduction (~70% compared to the value at ~5eV) at ~ 10eV, a broad peak around ~15eV and a dip again at ~ 22eV. These peaks in the backscattering cross sections combined with the electron capture probability may lead to the peak Ps formation at 5eV and another broad peak at around 20eV in our measurements.

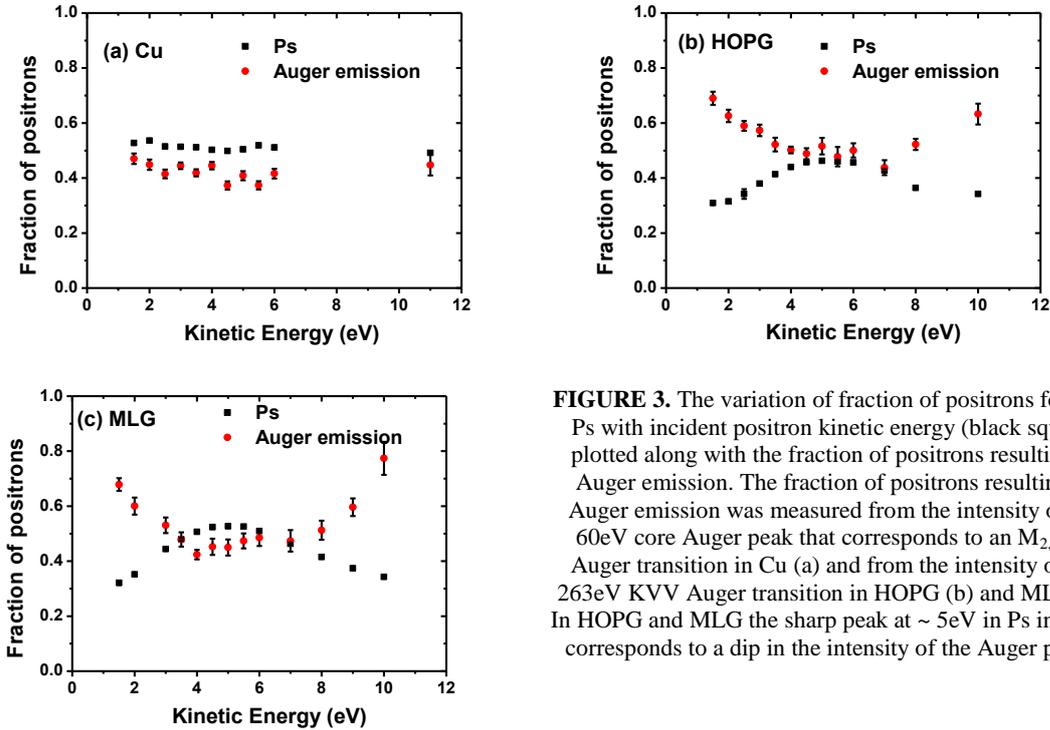

**FIGURE 3.** The variation of fraction of positrons forming Ps with incident positron kinetic energy (black square) plotted along with the fraction of positrons resulting in Auger emission. The fraction of positrons resulting in Auger emission was measured from the intensity of the 60eV core Auger peak that corresponds to an $M_{2,3}VV$ Auger transition in Cu (a) and from the intensity of the 263eV KVV Auger transition in HOPG (b) and MLG (c). In HOPG and MLG the sharp peak at ~ 5eV in Ps intensity corresponds to a dip in the intensity of the Auger peaks.

The second hypothesis is suggested by the fact that Ps formation can be considered to be similar to the neutralization of a positive ion and various processes of positive ion neutralization at sample surface should also be available for positrons. The major channels are Auger neutralization, where the excess energy released by the formation of Ps results in an electron emission, and resonant neutralisation where the electron tunnels from the sample to form Ps. The third channel is Ps formation through a mechanism by which the energy released through the binding of positron and electron results in the collective excitation of electrons at the surface. Ps formation by surface Plasmon excitation has been considered theoretically previously [14] and was shown to be a possible channel for Ps formation from metal surfaces especially at low positron energies. Graphite, a semi-metal, has two surface Plasmon peaks in the electron energy loss spectra, one corresponding to the π-Plasmon at ~ 6.5eV and another corresponding to (π+σ) Plasmon at ~ 25eV [15]. The minimum energy required for the incident positron to form Ps by Plasmon excitation is

$$E_{e^+} = E_{Plasmon} + \varphi^- - 6.8 \, eV \qquad (1)$$

where $E_{e^+}$ is the minimum kinetic energy of the positron, $E_{Plasmon}$ is the energy of the surface Plasmon mode that is to be excited during Ps formation, $\varphi^-$ is the sample work function and 6.8eV is the Ps binding energy. Thus, to excite the (π+σ) Plasmon, the incident positron energy should be greater than 22.7eV if the work function of graphite is assumed to be around 4.5eV. Below 22.7 eV, the positrons lose the (π+σ) Plasmon excitation channel for Ps formation which can explain the reduction in $f_{Ps}$ when the positron kinetic energy is less than 20eV. The broad nature of the Ps fraction near 20eV may reflect the broad nature of the (π+σ) Plasmon mode seen in electron energy loss spectra [15]. To excite the π Plasmon mode, the positron should have a minimum kinetic energy of 4.2eV and below this energy the surface Plasmon excitation channel is no longer available for positrons. This may explain the sharp feature around 5eV. The Plasmon mode is assumed to be one of the electron pick up processes during the backscattering process and hence the two peaks are super imposed on the increasing Ps formation with decreasing positron energy.

## CONCLUSION

Positronium (Ps) formation from the surfaces of highly oriented pyrolytic graphite (HOPG) and multi layer graphene (MLG) showed a sharp peak when the energy of the incident positrons was ~ 5eV with the Ps fraction being much higher than previously reported values for graphite. We propose that in HOPG and MLG, at low incident positron energies, the Ps formation is dominated either by a surface plasmon assisted electron pick up process or (and) by a back scattering process. Investigating the energy spectrum of the emitted Ps from HOPG and MLG as a function of incident positron energies may help to clarify on the origin of the spectral features observed in our measurements.

## ACKNOWLEDGMENTS

This work was supported by NSF grants DMR 1508719 and DMR 1338130, and Welch Foundation grant No. Y-1968-20180324. The authors thank Prof. B. Barbiellini for insightful discussions on Ps formation through Plasmon excitation.